\documentclass[preprint,showpacs,aps]{revtex4}
\usepackage[english]{babel}
\usepackage{graphicx,color}
\usepackage{latexsym}
\usepackage{amsmath}
\usepackage{amsfonts}
\usepackage{amssymb}
\usepackage[latin1]{inputenc}

\def\bb{\bibitem}

\def\aas{a\!\!\!/}
\def\bs{b\!\!/}

\def\ds{\partial\!\!\!/}

\def\bb{\bibitem}

\newcommand{\be}{\begin{equation}}
\newcommand{\ee}{\end{equation}}
\newcommand{\bea}{\begin{eqnarray}}
\newcommand{\eea}{\end{eqnarray}}

\newcommand{\I}{\mathrm{i}}
\newcommand{\E}{\mathrm{e}}
\newcommand{\D}{\mathrm{d}}

\begin{document}
\title{An Analogy of the Quantum Hall Condutivity in a Lorentz-symmetry Violation Setup}

\author{L. R. Ribeiro}
\email{lrr@fisica.ufpb.br}
\affiliation{{ Departamento de
F\'{\i}sica, Universidade Federal da Para\'\i ba, Caixa Postal 5008, 58051-970,
Jo\~ao Pessoa, PB, Brasil}}

\author{E. Passos}
\email{passos@fisica.ufpb.br}
\affiliation{Departamento de F\'\i sica,
Universidade Federal de Campina Grande, Caixa Postal 10071,
58109-970  Campina Grande, Para\'\i ba,
Brazil
}

\author{C. Furtado}
\email{furtado@fisica.ufpb.br}
\affiliation{{ Departamento de
F\'{\i}sica, Universidade Federal da Para\'\i ba, Caixa Postal 5008, 58051-970,
Jo\~ao Pessoa, PB, Brasil}}

\begin{abstract}We investigated some influences of unconventional physics, such Lorentz-symmetry violation, for quantum mechanical systems. In this context, we calculated a important contribution for Standard Model Extension. In the non-relativistic limit, we obtained a analogy of the Landau levels and the quantum Hall conductivity related to this contribution for low energy systems.
\end{abstract}

\maketitle

\section{Introduction}
Recent studies investigated the possibility of Lorentz-symmetry violation from the generalization of the usual Standard Model. This is called the Standard-Model Extension (SME), that has all the conventional properties but breaks the Lorentz-symmetry \cite{LV}.  In summary, this framework provides a quantitative description of Lorentz-symmetry violation, controlled by a set of coefficients whose the values are to be determined or constrained by experiments \cite{exp}. As an example, the QED sector of the minimal (SME) is given here. The QED extension
is obtained by restricting the minimal (SME) to the fermion and gauge field sectors. We rewrote the
fermion sector in the form
\bea\label{E1}
{\cal L}_{f}=\frac{i}{2}\bar\psi\Gamma^{\nu}D_{\nu}\psi-\bar\psi M \psi\;,
\eea
where the quantities $\Gamma^{\nu}$ and $M$ denote
\bea\label{E2}
\Gamma^{\nu}=\gamma^{\nu}+ c^{\mu\nu}\gamma_{\mu}+ d^{\mu\nu}\gamma_{5}\gamma_{\mu}\;,
\eea
and
\bea\label{E3}
M=m + \aas + \gamma_{5}\bs + \frac{1}{2} H^{\mu\nu}\sigma_{\mu\nu}\;.
\eea
The parameters $a, b, c, d$, and $H$ are fixed background expectation values of tensor fields that breaks conventional particle Lorentz-symmetry. The gauge field sector is
\bea\label{E4}
{\cal L}_{g}=-\frac{1}{4}F^{\mu\nu}F_{\mu\nu}+\frac{1}{2}\upsilon_{\mu}{^{*}F}^{\mu\nu}A_{\nu}+\cdots\;,
\eea
where $^*F^{\mu\nu}=\frac12\epsilon^{\mu\nu\rho\sigma}F_{\rho\sigma}$ is the usual dual field-strength electromagnetic tensor and $\upsilon_{\mu}$ is the fixed background tensor fields. Introduced by Carroll, Field and Jackiw \cite{Jac}, the  extension of the QED (\ref{E4}), does not break the
gauge symmetry of the action and equations of motion but it 
modifies the dispersion relations for different polarizations of
photons. Interesting investigations in the
context of Lorentz-symmetry violation have recently appeared  in the
literature, where several issues were addressed, such as
\v{C}erenkov-type mechanism called ``vacuum \v{C}erenkov radiation''
to test the Lorentz symmetry \cite{ptl}, changing of gravitational
redshifts for differently polarized Maxwell-Chern-Simons photons
\cite{kkl}, evidence for the Lorentz-CPT violation from the
measurement of CMB polarization \cite{bjx}, supersymmetric
extensions \cite{bch}, breaking of the Lorentz group down to the
little group associated with $\upsilon_\mu$ \cite{hle}, magnetic
monopoles inducing electric current \cite{brz} and radiative corrections argued \cite{cr}. 

In addition to the example of QED, we have new modifications to Standard Model of the type: 
\bea\label{E5}
{\cal L}_{new}=\bar\psi  b_{\lambda}{^{*}F}^{\lambda\rho}\gamma_{\rho}\psi\;.
\eea
This coupling describes the dynamics of a single neutral particle moving in the presence of the electromagnetic field and a constant parameter $b_{\lambda}$ which controls the Lorentz-symmetry violation. We can observe in the expression (\ref{E5}) that it is analog to a non-minimal coupling, and this background represents a adaptation of the Chern-Simons-like term (\ref{E4}) to Dirac equation. Interesting investigations on the non-relativistic content of this new background have also recently appeared  in the
literature, as non-minimal coupling and implications \cite{m1,m11,m4}, geometric phases \cite{k7,LECJ,m2}, Landau analog levels \cite{ELCJ}, magnetic moment generation \cite{Hl}.

Our interest is devoted to different features of new coupling found in (\ref{E5}).
The purpose of this paper is to study the possibility of calculate this coupling, and after analyzer unconventional implications on quantum Hall effect. The quantum Hall effect is one of the most interesting condensed-matter
phenomena discovered in the second half of the 20th century. This effect is observed in two dimensional electrons at very low temperatures and in strong magnetic fields, and its discover have fundamental significance as a manifestation of quantum mechanics on macroscopic scales \cite{klitz,tsui,laughlin}. Now, the quantum Hall effect is used to maintain the standard of electrical resistance by metrology laboratories around the world, since the fine structure constant is related to the quantum Hall conductivity. Nowadays, the quantum Hall effect is being study in the context of the unconventional physics, more specifically, at the non-commutative geometry \cite{nct,dayi}. 

Using the system of natural unit, the work structure is organized as follows: In section \ref{secI} we propose a certain model and through its, we calculate the coupling (\ref{E5}) via simple algebraic manipulation. In section \ref{secII} we coupling system (\ref{E5}) to Dirac equation e we determine the non-relativistic Hamiltonian associate using the Foldy-Wouthuysen (FW) transformation. In the section \ref{secIII} we calculate the Landau analog levels for space-like and time-like case independently. In the section \ref{secIV} we calculate the quantum Hall analog conductivity also for space-like and time-like case independently. Finally, in section \ref{secV}, we present our conclusions.

\section{Lorentz-symmetry violation coupling}\label{secI}
We wish now to study the possible appearance of relevant coupling found in the expression (\ref{E5}). For this purpose, we construct a interaction between the Dirac particle with usual field-strength electromagnetic tensor $F_{\mu\nu}$ and an constant parameter that controls the Lorentz-symmetry violation. Therefore, we start with the following Dirac equation modified:
\bea\label{03}
\big(i\ds- g b_{\lambda}{^{*}F}^{\lambda\rho}\gamma_{\rho}- m\big)\psi=0\;,
\eea
where the second term in (\ref{03}) is exactly the new modifications to (SME). This Lorentz-symmetry violation dependent contribution reads
\bea\label{04}
 g b_{\lambda}{^{*}F}^{\lambda\rho}\gamma_{\rho}=g b_{0}\vec{\gamma}\cdot\vec{B}+g\vec{b}\cdot\vec{B}\gamma_{0}- g\vec\gamma\cdot(\vec{b}\times\vec{E})
\eea
where $b_{0}$ and $\vec{b}$ are time-like and space-like component of the coefficient $b_{\mu}$ respectively. We have also that $\gamma^{0}$ is hermitian, $\vec\gamma$ is anti-hermitian, and are related to the $\hat\beta$ and $\vec\alpha$ matrices through: $\gamma^{0}=\hat\beta$, $\vec\gamma=\hat\beta\vec{\alpha}$.
We choose the Dirac matrices in the form,
\bea
\hat{\beta}=\gamma^{0}=\left(\begin{array}{cc}
1 & 0 \\
0 & -1\\ \end{array}\right)\;{\rm and}\,\,\vec{\gamma}=\left(\begin{array}{cc}
0 & \vec\sigma \\
-\vec\sigma & 0\\ \end{array}\right)\;.
\eea



\section{Non-relativistic limit}\label{secII}
At this moment, we are interested just in the contribution that introduces the Lorentz-symmetry violation in the system. Then, we rewrite the Eq.(\ref{03}) in the form
\bea\label{2.1}
\big(i\ds +g b_{0}\vec{\gamma}\cdot\vec{B}+g\vec{b}\cdot\vec{B}\gamma_{0}- g\vec\gamma\cdot(\vec{b}\times\vec{E})-m\big)\psi=0\;.
\eea 
This system describes a neutral spin half particle moving in a background that breaks the Lorentz-symmetry, which can be written as 
\be
i\partial_{0}\psi=H\psi=[\vec{\alpha}\cdot\vec{\pi}
+g\vec{b}\cdot\vec{B}+\hat{\beta}m]\psi\, ,
\ee
where $\vec{\pi}=-i(\vec{\nabla}+\hat{\beta}(ig\,(\vec{b}\times\vec{E}-\,b_{0}\vec{B}))$. At this point, we want to find the non-relativistic approach of our theory. Hence, we may use the Foldy--Wouthuysen (FW) transformation for the Dirac spinor \cite{fw}. According to the references \cite{LECJ} and \cite{ELCJ}, we may obtain the following Hamiltonian for this case:
\bea\label{eq:03}
\hat{H}\approx \hat{\beta}\Bigl[m-\frac{1}{2m}\Bigl(\vec{\nabla}+i\vec{A}\Bigl)^{2}+ A_{0}\Bigl]
\eea
with
\bea
\vec{A}=g \vec{b}\times\vec{E}-gb_{0}\vec{B}
\eea
and
\be
A_{0}=\frac{g\,\hat{\beta}}{2m}\vec{\Sigma}\cdot\big[\vec{\nabla}\times(\vec{b}\times\vec{E})\big]-\frac{g\hat{\beta}b_{0}}{2m}\vec{\Sigma}\cdot\big[\vec{\nabla}\times\vec{B}\big]
+g\vec{b}\cdot\vec{B}
\ee
The expression (\ref{eq:03}) is the non-relativistic quantum Hamiltonian for four-components fermi\-ons. However, for several applications in quantum mechanics, we can write (\ref{eq:03}) in the two-components fermions form
\be\label{ham}
H=- \frac{1}{2m}\Bigl(\vec{\nabla}+i\vec{a}\Bigl)^{2}+a_{0}\;,
\ee
that is similar to the interaction of a particle with the electric and magnetic fields minimally coupled to a non-Abelian gauge field with potential $a_\mu$, where
\be\label{eq:04}
a_{0}=\frac{g}{2m}\vec{\sigma}\cdot\big[\vec{\nabla}\times(\vec{b}\times\vec{E})\big]-\frac{gb_{0}}{2m}\vec{\sigma}\cdot\big[\vec{\nabla}\times\vec{B}\big]
+g\vec{b}\cdot\vec{B}\;,
\ee
and 
\begin{equation}
	\vec{a}=g \vec{b}\times\vec{E}-gb_{0}\vec{B}\;.
	\label{eq:04.1}
\end{equation} 
Here $\vec{\sigma}=(\sigma_1,\sigma_2,\sigma_3)$, $\sigma_i$ $(i=1,2,3)$ are the $2\times2$ Pauli matrices.
The Hamiltonian (\ref{ham}) describes a system formed by a neutral particle under influence of a constant parameter which controls the Lorentz-symmetry violation in the presence of electric and magnetic fields. 
\section{Landau levels}\label{secIII}
In 1930, Landau showed that a charged particle moving in an homogeneous magnetic field presents quantized energy levels \cite{landau}. The Landau levels have a remarkable role in the study of several problems in physics, e.g. quantum Hall effect \cite{prange}, different two-dimensional surfaces \cite{comtet,dunne}, anyons excitations in a rotating Bose--Einstein condensate \cite{paredes1,paredes2}, and others like analogies of Landau levels for dipoles. Ericsson and Sj\"oqvist developed a analogy of Landau quantization for neutral particles in presence of a external electric field \cite{ericsson}. The idea was based on the Aharonov--Casher effect in which neutral particles may interact with an electric field via a non-zero magnetic dipole moment. In the same way, we developed a analogy of Landau quantization for neutral particles, that possess a non-zero electric dipole moment, making use of the He--McKellar--Wilkens effect \cite{lrr:pla1}. To solve the problem of magnetic monopoles, we proposed the study of a analogy of Landau quantization in quantum dynamics of an induced electric dipole in the presence of crossed electric and magnetic fields \cite{lrr:pla2}. Several others Landau levels studies can be found in the literature \cite{k1, k2, k3, k4, k5, k6, m5}.

Here, we investigate a analogy of Landau quantization for a system of a neutral particle in the presence of a background that violates the Lorentz-symmetry. We separate the coupling as space-like and time-like cases, as follow.

\subsection{Space-like case}
 
Considering the space-like terms only in Hamiltonian (\ref{ham}), and taking the electric field at the $x$-$y$ plane\footnote{If we choose the electric field in the $x$-$y$ plane, the terms that contain the spin matrix $\sigma$ may simplified if we take the spinor up-component only.}\cite{mirza}, we have the space-like Hamiltonian as follow
\begin{equation}
H=\frac{1}{2m}\left[\vec{p}-g(\vec{b}\times\vec{E})\right]^2+\frac{ g}{2m}\,\vec{z}\cdot\bigl[\vec{\nabla}\times(\vec{b}\times\vec{E})\bigl]+g\,\vec{b}\cdot\vec{B}\;.
\label{eq:hamsl}
\end{equation}
We may obtain a analogy of Landau levels for this system under determined conditions \cite{ericsson}. These conditions are vanishing torque over the particle, static electric field $\partial_t\vec{E}=0$ and $\nabla\times\vec{E}=0$, and a uniform effective magnetic field. With the particle confined in the $x$-$y$ plane, these conditions are fulfilled by the following choices
\begin{eqnarray}
	\vec{z}&=&(0,0,1)\;,\nonumber\\
	\vec{E}&=&{{E_0}}(x,0,0)\;,\label{eq:fields}\\
	\vec{B}&=&B_0(0,0,x)\;,\nonumber
\end{eqnarray}
where ${{E_0}}$ and $B_0$ are constants. We define the following effective quantities
\begin{eqnarray}
	\vec{A}_\textrm{eff}&=&\vec{z}\times\vec{E}={{E_0}}(0,x,0)\;,\nonumber\\[-3.3mm]\label{eq:hc3}\\[-3.3mm]
	\vec{B}_\textrm{eff}&=&\nabla\times\vec{A}_\textrm{eff}={{E_0}}\vec{z}\;.\nonumber
\end{eqnarray}
Note that $\vec{B}_\textrm{eff}$ is homogeneous and satisfies one of the conditions for the analogy of Landau quantization occurs \cite{ericsson}. The term $g\vec{b}\cdot\vec{B}$ in the Eq. (\ref{eq:hamsl}) may be understood as a effective electric scalar potential, taking $\vec{b}=b_3\vec{z}$ we have
\begin{equation}
	V(x)=gb_3\vec{z}\cdot\vec{B}=gb_3B_0x\;,
	\label{eq:hc4}
\end{equation}
that generates a effective crossed electric field
\begin{equation}
	\vec{E}_\textrm{eff}=-\nabla V(x)=-gb_3B_0(1,0,0)\;,
	\label{eq:hc5}
\end{equation}
that is a constant field parallel to the plane where the particles can move. Now, we may rewrite the Hamiltonian (\ref{eq:hamsl}) in the form 
\begin{eqnarray}
	H=\frac{1}{2m}\big[p_x^2+(p_y-gb_3{{E_0}} x)^2\big]-\frac{gb_3{{E_0}}}{2m}+gb_3B_0x\;,
\end{eqnarray}	
or	
\begin{eqnarray}	H=\frac{p_x^2}{2m}+\frac12m\omega^2\left[x-\left(p_y\ell^2-\frac{gb_3B_0}{m\omega^2}\right)\right]^2-\frac{\omega}{2}+gb_3B_0\left(p_y\ell^2-\frac{gb_3 B_0}{m\omega^2}\right)\;,
	\label{eq:hc6}
\end{eqnarray}
where $\omega=\frac{gb_3{{E_0}}}{m}$ is the cyclotron frequency and $\ell=\frac{1}{\sqrt{gb_3{{E_0}}}}$ is the fundamental length. Since the Hamiltonian (\ref{eq:hc6}) does not explicitly depend  on $y$, we may use the ansatz $\Psi(x)=\exp{(\I ky)}\Phi(x)$ to the wave function, where $k$ is the eigenvalue of $p_y$. We define $X_k=\left(k\ell^2-\frac{gb_3B_0}{m\omega^2}\right)$ as the center of harmonic oscillator and rewrite the Hamiltonian (\ref{eq:hc6}) in the form
\begin{equation}	H=\frac{p_x^2}{2m}+\frac12m\omega^2\left(x-X_k\right)^2-\frac{\omega}{2}+gb_3B_0X_k+\frac12mv_\textrm{D}^2\;,
	\label{eq:hc7}
\end{equation}
where $v_\textrm{D}=\dfrac{B_0}{{{E_0}}}$ is related to the drift velocity $\dfrac{\vec{E}_\textrm{eff}\times\vec{B}_\textrm{eff}}{|\vec{B}_\textrm{eff}|^2}=gb_3\dfrac{B_0}{{{E_0}}}\vec{y}$. 
The energy eigenvalues are written as
\begin{equation}	\mathcal{E}_{nk}=\left[n+\frac{1-\varsigma}{2}\right]|\omega|+gb_3B_0X_k+\frac{1}{2}mv_\textrm{D}^2\;,
	\label{eq:hc8}
\end{equation}
where $\varsigma=\pm$ labels the revolution direction of the cyclotron frequency $\omega=\varsigma|\omega|$. Note that the energy eigenvalues are linearly dependent on the position of the center of the oscillator $X_k$, and therefore linear in the momentum component $p_y$. Solving the harmonic oscillator differential Eq.(\ref{eq:hc7}), we found the eigenfunction to the lowest Landau level
\begin{equation}
	\Phi_0(x)=\frac{1}{\sqrt{\ell\sqrt{\pi}}}\E^{-\frac{1}{2\ell^2}(x-X_k)^2}\;.
	\label{eq:hc9}
\end{equation}
Higher levels are related to Hermite functions, but it is not necessary to our goals.

\subsection{Time-like case}
In the same way of the space-like case, if we consider the time-like part of Hamiltonian (\ref{ham}), also considering the fields in the $x$-$y$ plane, we have
\begin{equation}
	H=\frac{1}{2m}\left(\vec{p}+g b_{0}\vec{B}\right)^2-\frac{ g b_{0}}{2m}\,\vec{z}\cdot\bigl[\vec{\nabla}\times\vec{B}\bigl]\;.
	\label{eq:hc14}
\end{equation} 
In the same way of the space-like case, we may choose a field configuration in which a analogy of Landau quantization occurs,
\begin{eqnarray}
  \vec{z}&=&(0,0,1)\;,\nonumber\\
	\vec{A}_\textrm{eff}&=&\vec{B}=B_0(0,x,0)\;,\\
	\vec{B}_\textrm{eff}&=&\nabla\times\vec{A}_\textrm{eff}=B_0\vec{z}\nonumber\;,
	\label{eq:hc15}
\end{eqnarray}
where $B_0$ is a constant. Thus, we rewrite the Hamiltonian (\ref{eq:hc14}) in the form
\begin{equation}
	H=\frac{p_x^2}{2m}+\frac12m\omega^2\left(x-k\ell^2\right)^2+\frac{\omega}{2}\;,
	\label{eq:hc16}
\end{equation}
where we used the ansatz $\Psi=\E^{\I ky}\Phi(x)$, $k$ is the eigenvalue of $p_y$ and $\ell=\frac{1}{\sqrt{gb_0B_0}}$ is the fundamental length. Here, $\omega=\frac{gb_0B_0}{m}$ is the cyclotron frequency and $\omega=\varsigma|\omega|$ where $\varsigma=\pm$ labels the signal of cyclotron revolution. The Eq.(\ref{eq:hc16}) is a harmonic oscillator Hamiltonian, and have quantized energy eigenvalues
\begin{equation}
	\mathcal{E}_{n}=\left[n+\frac{1+\varsigma}{2}\right]|\omega|\;.
	\label{eq:hc17}
\end{equation}
The Landau lowest levels can be calculated
\begin{equation}
	\Phi_0(x)=\frac{1}{\sqrt{\ell\sqrt{\pi}}}\E^{-\frac{1}{2\ell^2}(x-k\ell^2)^2}\;.
	\label{eq:hc18}
\end{equation}
In this case, the energy do not depend on the certer of harmonic oscilator.

\section{The Hall conductivity}\label{secIV}
Since we have the lowest Landau levels, we may calculate the average value of particles current in the $x$-$y$ plane, and find a analogy of the Hall conductivity for our system. If we consider the strong fields limit, the spacing between Landau levels becomes infinite, and the lowest Landau level decouples form all the higher levels. Thus, this limit projects the quantum mechanical system onto the lowest Landau level. Considering the system of a neutral particle in the presence of a background that violates the Lorentz-symmetry, we analyze the space-like and time-like cases as follow. 

\subsection{Space-like case}

Assuming the effective magnetic field very large, so we can ignore higher Landau levels, and take the lowest level only in (\ref{eq:hc9}). Thus, we calculate the expectation value of current
\begin{equation}
	\langle\vec{J}\rangle=-\frac{gb_3\varrho}{m}\langle\Psi_0|\vec{p}-gb_3\vec{A}_\textrm{eff}|\Psi_0\rangle\;,\label{eq:hc10}
\end{equation}
that vanishes to the $x$ component
\begin{equation}
	\langle J_x\rangle=-\frac{gb_3\varrho}{m}\langle\Psi_0|p_x|\Psi_0\rangle=0\;,\label{eq:hc11a}
\end{equation}
and to the $y$ component results
\begin{eqnarray}
	\langle J_y\rangle&=&-\frac{gb_3\varrho}{m}\langle\Psi_0|p_y-gb_3{{E_0}} x|\Psi_0\rangle\nonumber\\
	&=&-\frac{gb_3\varrho}{m}\frac{1}{\ell\sqrt{\pi}}\int{\E^{-\frac{1}{\ell^2}(x-X_k)^2}(k-gb_3{{E_0}} x)}\,\D x\nonumber\\
	&=&-\frac{gb_3\varrho}{m}(k-gb_3{{E_0}} X_k)=-gb_3\varrho\frac{B_0}{{{E_0}}}\nonumber\\&=&-gb_3\varrho v_\textrm{D}\label{eq:hc11}\;,
\end{eqnarray}
where $\varrho$ is the density of particles. Now, we can find the conductivity
\begin{equation}
	\left(\begin{array}{c}
	J_x\\J_y
	\end{array}\right)=
	\left(\begin{array}{cc}
	\sigma_{xx}&\sigma_{xy}\\
	-\sigma_{xy}&\sigma_{xx}
	\end{array}\right)
	\left(\begin{array}{c}
	E_x\\E_y
	\end{array}\right)\;,
	\label{eq:hc12}
\end{equation}
where $\sigma_{ij}$ is the matrix formed by the elements of the conductivity tensor in the $xy$-plane, and $\vec{E}_\textrm{eff}=(E_x,E_y)$, so $E_x=-gb_3B_0$ and $E_y=0$. Thus, the Hall conductivity $\sigma_\textrm{H}=\sigma_{xy}$ is found
\begin{equation}
	\sigma_\textrm{H}=-\nu gb_3\;,\qquad\nu=\frac{\varrho\phi_0}{E_0}\;,
	\label{eq:hc13}
\end{equation}
where $\phi_0=(gb_3)^{-1}$ is the effective flux quantum. The Landau level filling factor $\nu=\frac{\varrho}{(E_0/\phi_0)}$ is the ratio of particles density $\varrho$ and the effective magnetic flux density $\frac{E_0}{\phi_0}$; thus $\nu$ is about $\frac{p}{q}$ ($p$ and $q$ integer) and may be a integer or fractional number.

\subsection{Time-like case}

Also supposing the effective magnetic field very large, so that we can ignore higher Landau levels, and take the lowest level only (\ref{eq:hc18}). Thus, we calculate the expectation value of current in the $x$-$y$ plane. Again, the $x$ component of current vanishes
\begin{equation}
	\langle J_x\rangle=-\frac{gb_0\varrho}{m}\langle\Psi_0|p_x|\Psi_0\rangle=0\;,
	\label{eq:hc19a}
\end{equation}
and the $y$ component vanishes too
\begin{eqnarray}
	\langle J_y\rangle&=&-\frac{gb_0\varrho}{m}\langle\Psi_0|p_y-gb_0{{B_0}} x|\Psi_0\rangle\nonumber\\
	&=&-\frac{gb_0\varrho}{m}\frac{1}{\ell\sqrt{\pi}}\int{\E^{-\frac{1}{\ell^2}(x-k\ell^2)^2}(k-gb_0{{B_0}} x)}\,\D x\nonumber\\
	&=&-\frac{gb_0\varrho}{m}(k-gb_0{{B_0}}k\ell^2)=0\;.\label{eq:hc19}
\end{eqnarray}
Because the absence of the crossed effective electric field, the current density is antisymmetric about the
peak of the gaussian and hence the total current vanishes. Therefore, we have no Hall conductivity for the time-like case.


\section{Conclusions}
\label{secV}
We investigate different situations associated to a certain coupling which can modify the
fermionic sector of Standard Model Lagrangian. First, we use a convenient fermionic sector, composed by a term suggested by (SME), and a usual coupling that describes a neutral particle anomalous moment, and we show that such coupling may be calculated via algebraic manipulation. This result is very important because now we know that such modification in Lorentz-symmetry breaking can be obtained from other model previously well-known without needing to use a procedure more complicate such as radiative correction.

We write a Dirac equation modified by the coupling cited above and, at the non-relativistic limit, we construct a analogy of the Landau levels for a neutral particle system. We analyze the space-like and time-like cases, and find the energy eigenvalues and fundamental state eigenfunctions for both cases. At strong fields, the lowest Landau level decouples from the all of rest, and this limit projects the system onto lowest Landau level.

With the advent of a analogy of Landau levels, we calculate the average current and the Hall conductivity in the space-like case. We write the Hall conductivity in terms of the Landau filling factors, that may be a integer or fractional number. In the time-like case, we have no conductivity at the particles moving plane due the absence of a external scalar potential which generates the effective crossed electric field. Thus, the current density is antisymmetric about the peak of the probability density gaussian and the current expectation value vanishes, as well as the conductivity tensor in this case. It's well know that Hall conductivity can be experimentally measured with very high accuracy. Thus it opens good opportunities to detect very small fenomena like Lorentz-symmetry breaking ones.

{\bf Acknowledgments.} The authors are grateful to J. R. Nascimento for some criticism of the manuscript. This work was partially supported by PRONEX/FAPESQ-PB, FINEP, CNPq and CAPES/PROCAD.

\end{document}